\documentclass[12pt]{article}
\usepackage{amssymb,amscd}
\usepackage{epsfig}
\begin{document}

\begin{center}
{\bf CHERENKOV GLUE IN OPAQUE NUCLEAR MEDIUM}
\end{center}

\bigskip

\begin{center}
\bf{ \large I.M Dremin$^{(a)}$\footnote{Supported by the RFBR grants 09-02-00741
and 08-02-91000-CERN},
M.R. Kirakosyan$^{(a)}$\footnote{Supported by the RFBR grants 09-02-00741
and 08-02-91000-CERN}, \\
A.V. Leonidov$^{(a,b)}$\footnote{Supported by the RFBR grants 09-02-00741
and 08-02-91000-CERN},
A.V. Vinogradov$^{(a)}$\footnote{Supported by the RFBR grant 08-02-91000-CERN}}
\end{center}
\medskip

(a) {\it  P.N.~Lebedev Physics Institute, Moscow, Russia}

(b) {\it Institute of Theoretical and Experimental Physics, Moscow, Russia}



\bigskip

\begin{abstract}
The spectrum of Cherenkov gluons is calculated within the framework of in-medium QCD. It is compared with experimental data on the double-humped
structure around the away-side jet obtained in the experiments STAR and PHENIX at RHIC. The values of the real and imaginary parts of the
permittivity of the medium created in ultrarelativistic nuclear collisions are obtained from these fits. It is shown that accounting for an
additional smearing of transverse momenta of Cherenkov gluons in the medium allows to provide a very good description of the experimental data.
\end{abstract}

\newpage

Remarkable experimental results revealing an existence of double-humped structure of away-side azimuthal correlations in central high energy
nuclear collisions obtained in the experiments STAR \cite{STAR05,U08} and PHENIX \cite{PHENIX06,PHENIX08a,PHENIX08b} at RHIC  have triggered a new
wave of interest to a phenomenon of Cherenkov radiation of gluons first considered in \cite{D79,D81,ADDK79}. A simple model of scalar excitations
in strongly interacting quark-gluon plasma leading to Cherenkov-like pattern was considered in \cite{KMW06}. An alternative explanation of the
observed double - humped structure in terms of Mach cones was discussed in \cite{MR05,NMR08}. For a recent comparison of these two mechanisms see
\cite{D08}. Let us note however that a recent detailed study \cite{BNTGMR08} presents strong arguments against the Mach-cone interpretation by
proving that a dominating feature of hydrodynamical picture of energy loss is a strong diffusion wake leading to a pronounced maximum in the
forward direction.

In an ideal case of transparent medium Cherenkov gluons are radiated at fixed angle with respect to the direction of propagation of the color
current. Their spectrum is described by the well-known Tamm-Frank formula \cite{TF37,T39}. The experimentally observed double-humped angular
distribution shows the structure that is noticeably spread around the angles corresponding to the maxima. Among the physical reasons causing such
spreading are, in particular, the opacity of the color medium and the finite propagation length of the color current. Some effects were treated in
\cite{IJMP}. In the present paper we consider a model combining an account for the opacity of the color medium  and an additional smearing of
"pure" Cherenkov pattern in the plane orthogonal to a propagation direction of the radiating current that allows to achieve a good description of
the available experimental data.

The equations of in-medium QCD were proposed in \cite{MQCD}. At tree level and in quasi-abelian approximation they coincide (up to color factors)
with the corresponding equations in QED. Therefore, in this approximation Cherenkov gluons are described in complete analogy with Cherenkov
photons and correspond to the specific singularity of an abelian Lienard-Wiechert potential having its origin in the nontrivial (chromo)dielectric
permittivity $\epsilon$. In \cite{MQCD} the simplest case of color singlet constant real permittivity was considered. In the present paper, still
working in the quasi-abelian approximation and considering the color singlet permittivity, we shall consider a more general case of opaque medium
characterized by complex permittivity
\begin{equation}
 \epsilon(\omega ) \, = \, \epsilon_1(\omega ) + i \, \epsilon_2(\omega ),
\end{equation}
where $\epsilon_1(\omega )=\epsilon_1(-\omega )$, $\epsilon_2(\omega )= -\epsilon_2(-\omega )$, $\omega $ is the gluon energy.  For simplicity we
neglect the spatial dispersion.

Truly non-abelian effects can be considered by taking into account the nontrivial tensor chromoelectric permittivity in the color space and the
nonlinear terms in the Yang-Mills equations in the medium \cite{MQCD}. These effects will be considered in the future publication \cite{adl}.

To develop a description of the angular pattern observed in the flow of final hadrons in high energy nuclear collisions
\cite{STAR05,U08,PHENIX06,PHENIX08a,PHENIX08b} in terms of the Cherenkov gluon radiation one has to

{\bf (a)} Describe a kinematical pattern characterizing the initially produced hard partons serving as colored sources of gluon Cherenkov
radiation taking into account the experimental cuts on pseudorapidity and transverse momentum \cite{U08,PHENIX08b}.

{\bf (b)} Write down a spectrum of Cherenkov gluon radiation in the opaque medium for the above-described color sources.

{\bf (c)} Describe a conversion of Cherenkov gluons into observed hadrons taking into account the experimental cuts on transverse momenta of final
hadrons \cite{U08,PHENIX08b}.

These three steps are realized in a Monte-Carlo procedure. In total the model includes three parameters, described in more details below - two
related to the Cherenkov gluon radiation, the magnitudes of the real $\epsilon_1$ and imaginary $\epsilon_2$ parts of the permittivity of the
medium, and a scale $\Delta_\perp$ controlling the transverse smearing of Cherenkov gluons . The values of these parameters are fitted to achieve
the best possible agreement with the experimental spectra \cite{U08,PHENIX08b}.

The description of the initial hard color sources was performed through a Monte Carlo simulation of hard parton-parton scattering at RHIC energies
with PYTHIA \cite{PYTHIA}. An initial pool of two-parton configurations resulting from these scatterings consisted of those in which the trigger
jet gave rise to a pion hitting the trigger intervals in rapidity, $|\,\eta|^{\rm \, tr}_{\rm \, star} < 0.7$ and $|\,\eta|^{\rm \, tr}_{\rm \,
phenix} < 0.35$, and transverse momentum, where we have chosen the intervals $3 \, {\rm GeV} < |\, {\bf p_\perp^{\rm lab}}| < 4 \, {\rm GeV}$ for
STAR \cite{U08} and $2 \, {\rm GeV} < |\, {\bf p_\perp^{\rm lab}}| < 3 \, {\rm GeV}$ for PHENIX \cite{PHENIX08b}. The fragmentation of gluon
(quark) generating the trigger near-side jet into pions was described by standard fragmentation functions measured at LEP \cite{A00}. For
simplicity we have considered only the dominating subset of events with gluonic away-side jet.

Let us now turn to the differential Cherenkov energy loss spectrum in the opaque medium per length $dl$ along the direction of the radiating
gluons (quarks) \cite{G02,GS03}
\begin{equation}\label{cherloss0}
 \frac{1}{\omega } \frac{d W}{dl \,d\omega \, d \phi \, d \cos \theta} =
 \frac{4 \alpha_S C_{V(F)}}{\pi} \frac{\cos \theta (1-\cos^2\theta)
 \Gamma(\omega )}{\left(\cos^2\theta-\zeta(\omega )\right)^2+\Gamma^2(\omega )},
\end{equation}
where $C_{V(F)}$ are the Casimir operators for the adjoint and fundamental representations of the color group correspondingly, $\theta , \phi $
are polar and azimuthal angles with respect to the emitter propagation direction and
\begin{eqnarray}
 \zeta(\omega ) & = & \frac{\epsilon_1(\omega )}{\epsilon_1^2(\omega )+
 \epsilon_2^2(\omega )}, \nonumber \\
 \Gamma(\omega ) & = & \frac{\epsilon_2(\omega )}{\epsilon_1^2(\omega )+
 \epsilon_2^2(\omega )}, \nonumber \\
\end{eqnarray}
so that $\zeta(\omega )$ controls the location of the maximum and $\Gamma(\omega )$
controls the spreading around it. This a'la Breit-Wigner shape replaces
the $\delta $-functional angular dependence characteristic for real
$\epsilon $. A derivation
of Eq.~(\ref{cherloss0}) in terms of the energy flow of radiated gluons
is given in Appendix 1. For $(\epsilon _2/\epsilon _1)^2\ll 1$, the
energy-dependent maximum of the differential spectrum (\ref{cherloss0}) is at
\begin{equation}\label{cosmax}
 \cos^2 \theta_{\rm max} (\omega ) \, \approx \, \frac{\epsilon_1(\omega )}{\epsilon_1^2(\omega )+\epsilon_2^2(\omega )}
 \, \approx \, \frac{1}{\epsilon_1(\omega)}.
\end{equation}
The second relation is just the Tamm-Frank expression \cite{TF37,T39} for ultrarelativistic particles. Let us stress that in Eq.~(\ref{cherloss0})
the angle $\theta$ of emission of Cherenkov gluons is measured with respect to the direction of propagation of the radiating color current which
in the considered case is that of an initial gluon of the away-side jet. In the ensemble of initial jets the near-side and away-side jets are in
general not angularly balanced in the laboratory system. However, it was argued in \cite{IJMP} that an angularly balanced configuration is in fact
dominant because a generic shape of structure functions leads to a suppression of energy-mismatched configurations of the initial partons and
consequently of the angular mismatched configurations of the final ones. In our Monte-Carlo program this mismatch is, nevertheless, taken into
account by using PYTHIA.

At the second step of our Monte-Carlo procedure we generate Cherenkov gluon radiation by the away-side gluon jet. To illustrate the corresponding
kinematics  let us consider a simple example of the Cherenkov spectrum generated by the source having a direction of propagation orthogonal to the
collision axis $z$ that allows an analytical treatment. In order to compare the spectrum (\ref{cherloss0}) with the experimental one in this
kinematics we should rewrite it in terms of experimental laboratory polar and azimuthal angles $\theta_L$ and $\phi_L$. It is easy to see that for
the above geometry
$$
    \cos \theta \, = \, \vert \sin \theta_L \cos \phi_L\vert ,
$$
for $\phi_L$ counted from the away-side jet.

The number $dN$ of emitted gluons per length $dl$ with energy and angles within intervals $d\omega \, d\theta_L \, d\phi_L $ is
\begin{equation}\label{speclab}
 \frac{dN}{dl \, d\omega \, d \phi_L \, d \cos \theta_L} =\frac{4 \alpha_SC_V}{\pi}
 \frac{\vert \sin \theta_L \cos \phi_L \vert (1-\sin^2 \theta_L \cos^2 \phi_L)
 \Gamma(\omega )}{\left(\sin^2 \theta_L\cos^2\phi_L-\zeta(\omega )\right)^2+
 \Gamma^2(\omega )}.
\end{equation}
The experimentally observed hadron spectrum results from hadronization  of the gluon distribution $dN/d\phi_L$ obtained from Eq.~(\ref{speclab})
by integrating over $\theta_L$:
\begin{equation}\label{cherloss}
 \frac{d N}{d\omega \, dl \, d \phi_L}= 4 \alpha_s C_V \frac{\Gamma}
 {\vert \cos\phi_L \vert} \left \{
 \frac{1}{\sqrt{\Gamma^2+\left( \cos^2 \phi_L - \zeta \right)^2} }
 \frac{1}{\sqrt{A^2+B^2}} \, \right. \times
\end{equation}
$$
 \left[
 \left(  A + (1-\zeta)\frac{B}{\Gamma} \right) \sqrt{\frac{1}{2}
 \left(\sqrt{A^2+B^2}+A \right)} \right. -
$$
$$
 \left. \left. \left( B + (1-\zeta)\frac{A}{\Gamma} \right) \sqrt{ \frac{1}{2}
 \left(\sqrt{A^2+B^2}-A \right)} \right] - 1 \right \},
$$
 where
\begin{eqnarray}
 A(\omega ,\phi_L) & = & \Gamma^2(\omega )+\zeta ^2(\omega )-\zeta (\omega )
 \cos^2 \phi_L ,\nonumber \\
 B(\omega ,\phi_L) & = & \Gamma(\omega ) \cos^2 \phi_L .
\end{eqnarray}

The gluon spectrum (\ref{cherloss}) reveals the double-humped structure of interest.

Let us stress once again that in our Monte-Carlo procedure we consider a more general situation where the direction of the gluon generating the
away-side jet is fixed within each configuration satisfying the above-described
trigger conditions imposed on the properties of the near-side jet in experiment.

To realize a  Monte-Carlo procedure of generating the Cherenkov spectrum we have to specify the functions $\epsilon_1(\omega)$ and
$\epsilon_2(\omega)$. Taking into account that a range of (laboratory frame) transverse momenta of the away-side hadrons in which the
double-humped structure is observed is quite narrow ( $\sim 2 \, {\rm GeV}$) and the binning of experimental data we aim to reproduce was
performed at a scale of $1 \, {\rm GeV}$ we can with a good accuracy neglect the effects of dispersion and consider energy-independent
$\epsilon_{1,2}={\rm const}$. The values of these constants are determined through fitting the experimental data. Within this assumption the
spectrum of produced Cherenkov glue is simply energy-independent, $d N / d\omega \, dl={\rm const}$. Realistically the possibility of Cherenkov
emission is restricted to some finite interval of energies $\omega < \omega_{\rm max}$ so that
\begin{equation}
\frac{dN}{d \omega \, dl} \propto \theta \left(\omega_{\rm max}-\omega \right)\, ,
\end{equation}
where $\omega_{\rm max}$ is the highest characteristic resonance excitation energy of the medium\footnote{For a discussion of interrelation
between Cherenkov gluons and hadronic resonances in the medium see \cite{dnec}.}. In our simulations we have chosen $\omega_{\rm max}=3.5\,{\rm
GeV}$. We have verified that our results are in fact weakly sensitive to the exact value of $\omega_{\rm max}$.

The thus generated flow of Cherenkov glue should, of course, be transformed into that of final hadrons. There exist several phenomenological
schemes describing this conversion. In our case it is convenient to use a language of fragmentation functions $D^h_g(x,{\bf p_\perp}|\,Q^2)$ which
generically describe a probability for a gluon with energy $E$ and virtuality $Q^2$ to produce a hadron with the energy $xE$ and transverse
momentum ${\bf p_\perp}$ measured with respect to the direction of propagation of the initial gluon.

One is first tempted, relying on the soft blanching hypothesis, to ascribe the same shape to the spectrum of hadrons resulting from fragmentation
of the Cherenkov hadrons which amounts to assuming $D^h_g(x,{\bf p_\perp}|\,Q^2) \propto \delta(1-x)$ as supported by the experimental evidence
obtained in the physics of multiparticle production from $e^+e^-\rightarrow $ jets. However, this does not lead to the fully satisfactory
description of experimental data because of predicting, in contradiction with the experimental data, that the probability of radiating Cherenkov
glue at angles $\vert \, \phi _L\vert \geq \pi/2$ with respect to the away-side jet is strictly zero. These "dead zones" appear because the
Cherenkov gluon "halo" can not spread to angles larger than $\pi /2$ to the direction of the emitter. It is clear that such "dead zones" will in
fact be present for any fragmentation mechanism that does not generate transverse momentum with respect to the direction of propagation of the
initial Cherenkov gluon which in our case corresponds to considering a ${\bf p_\perp}$-independent fragmentation function $D^h_g(x,|\,Q^2)$. We
confine our consideration to fragmentation to light hadrons because experimentally the share of other bosonic resonances is quite small (in
particular, $\rho :\omega :\phi $=10:1:2 according to \cite{damj}). In our analysis we used a simple parametrization of the fragmentation function
$D^h_g(x,{\bf p_\perp}|\,Q^2)$ {\it including} the effects of transverse smearing:
\begin{equation}
 D^h_g(x,{\bf p_\perp}|\,Q^2) \propto (1-x)^3 \frac{1}{\sqrt{2 \pi \Delta_\perp^2}} \exp \left\{ -\frac{\bf p_\perp^2}{2 \Delta_\perp^2}
\right\}\, ,
\end{equation}
where the $x$-dependence is that of a gluon fragmentation function at the reference scale $Q_0=2\,{\rm GeV}$ \cite{A00}.

Thus, we have a set of three parameters: $\epsilon_1, \epsilon_2$ and the transverse smearing scale  $\Delta_\perp$.

The values of the parameters that provide the best description of experimental data of STAR and PHENIX collaborations for the ranges of transverse
momenta in the associated jet $1 \, {\rm GeV} < |\, {\bf p_\perp^{\rm lab}}| < 2 \, {\rm GeV}$ for STAR and $2 \, {\rm GeV} < |\, {\bf
p_\perp^{\rm lab}}| < 3 \, {\rm GeV}$ for PHENIX are shown in Table 1.

\bigskip

\begin{center}
{\bf Table 1}

\bigskip

\begin{tabular}{|c|c|c|c|c|}
  \hline
  Experiment & $\theta_{\rm max}$ & $\epsilon_1$ & $\epsilon_2$ & $\Delta_\perp$ \\ \hline
  STAR &  1.04~rad &  5.4 & 0.7 & 0.7~{\rm GeV}\\ \hline
  PHENIX & 1.27~rad & 9.0 & 2.0 & 1.1~{\rm GeV}\\ \hline
\end{tabular}
\end{center}

\bigskip

Let us stress that a  difference in the fitted values of $\epsilon_{1}$ for STAR and PHENIX originates from different positions of angular maxima
$\theta_{\rm max}$ in the corresponding experimental data. (May be, it could indicate on necessity to take the dispersion into account.) However
these values are quite stable in both above approaches because they are defined by maxima positions but not by the humps widths.

The value of $\epsilon_{2}$, on the contrary, is influenced by the widths. Nevertheless, the ratio $(\epsilon_{2}/\epsilon_{1})^2 \lesssim 0.05$
stays quite small in all cases.

As to the transverse smearing parametrized by $\Delta_\perp$, one can get a feeling of the magnitude of this effect by considering a typical
distance covered by the away-side parton $L \simeq 5\,{\rm fm}$ \cite{WHDG07}. This gives a smearing rate per fermi of $\Delta_\perp^2 /fm \sim
0.2 \left[{\rm GeV}^2/{\rm fm} \right]$.

The resulting angular spectra for STAR and PHENIX are shown in Figs. 1 and 2 correspondingly.
\begin{figure}[ht]
 \epsfig{file=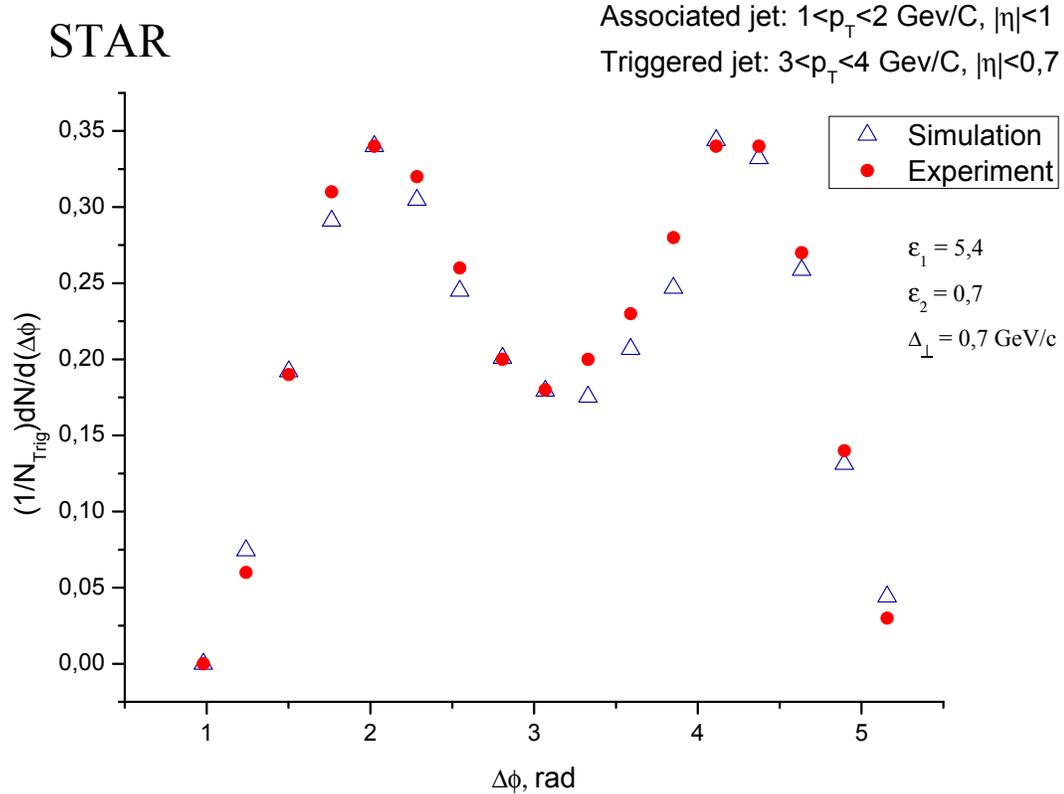,width=16cm}
 \caption{Away-side azimuthal correlations for STAR, circles - experimental data, triangles - simulation.}
\end{figure}

\begin{figure}[ht]
 \epsfig{file=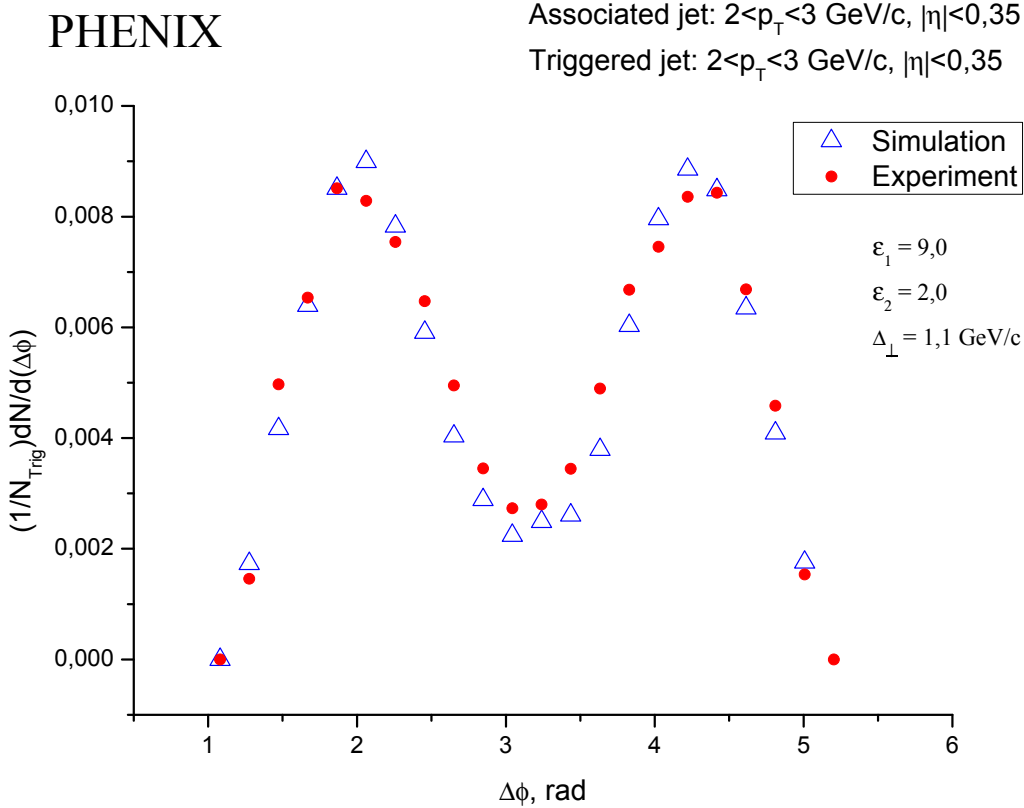,width=16cm}
 \caption{Away-side azimuthal correlations for PHENIX, circles - experimental data, triangles - simulation.}
\end{figure}
We see that the positions of the maxima (and therefore the values of $\epsilon_{1}$) are quite stable to accounting for additional smearing on top
of that in Eq.~(\ref{cherloss}). However, it is indeed important in achieving a good description of the widths of humps in experimental data as
seen in Figs. 1 and 2. The shape of humps in the former "dead zone" determines the parameter $\Delta_\perp $, which, in its turn, influences
$\epsilon _2$.

To conclude, the values of the real part of the nuclear permittivity $\epsilon_{1}$ found from the fit to experimental data of RHIC (albeit
somewhat different in the two experimental sets) are determined with good precision. Their common feature is that they are noticeably larger than
1. This shows that the density of scattering centers is quite large and the nuclear medium reminds a liquid rather than a gas (for more details
see \cite{IJMP}). The accuracy of the estimate of the values of $\epsilon_{2}$ is much less but more important is the fact that they are rather
small compared with $\epsilon_{1}$.

In this paper we have described a qualitative model that allows, in a framework of a reasonable phenomenological picture, to reproduce the
experimentally observed double-humped structure of the away-side jets in heavy ion collisions at RHIC
\cite{STAR05,U08,PHENIX06,PHENIX08a,PHENIX08b}. To develop a quantitative description one needs to consider a more specific microscopic model for
chromodielectric permittivity and study the effects of hadronization in more detail using Monte-Carlo simulations. Such studies aimed at working
out a quantitative explanation of the dependence of the effect on the range of triggered transverse momenta, in particular a disappearance of the
effect at increasing $p_\perp$, and a detailed description of the central domain related to fragmentation of the radiating particle are now in
progress. Let us stress that in working out such quantitative description it is crucially important to have experimental data in a broader range
of (laboratory) transverse momenta, in particular on configurations with a pronounced mismatch between the momenta of trigger and associated jets.

\bigskip

{\bf\large Acknowledgments }\\

I.D. thanks A.S.~ Sakharov and E.K.~Sarkisyan-Grinbaum for participation in this work at the earlier stage.

We are grateful to A.B. Kaidalov, A.D. Mironov and Yu.A. Simonov for useful discussion and J.G. Ulery for helpful communication.

The work of I.D., A.L. and M.K. was supported by RFBR grants 09-02-00741 and
08-02-91000-CERN. The work of A.V. was supported by RFBR grant 08-02-91000-CERN.\\

\bigskip

{\bf \large Appendix 1}

\bigskip

The energy losses due to emission of transverse gluons read
\begin{equation}
  \frac{dW}{dz}  =\frac{i C_V g^2}{2 \pi^2 }  \int d\omega \int d {\bf k}
  \,\, \frac{\omega }{{\bf k}^2} \,
  \frac{{\bf k}^2 v^2 -({\bf k} {\bf v})^2}{ {\bf k}^2 - \omega ^2
  \epsilon (\omega ) }
  \delta (\omega  - {\bf k}{\bf v}),
\end{equation}
where ${\bf v}$ is a unit vector determining the direction of propagation of the radiating color current. In cylindrical coordinates $(k,{\bf
q},\phi)$ with ${\bf v}=(v_x,v_y,v_z)=(0,0,1)$ we have
\begin{equation}
  \frac{d W}{dz}   =\frac{i C_V g^2}{\pi }  \int d\omega \omega
  \int_0^\infty dq \, q^3  \int_{-\infty}^\infty dk \frac{1}{k^2+q^2}
  \frac{1}{k^2+q^2-\epsilon (\omega ) \omega ^2} \delta (k-\omega ).
\end{equation}
An angle at which gluons are emitted can be found from the ratio
of the corresponding components of the Poynting
vector
\begin{equation}
 \tan^2 \theta = \frac{S^2_z}{S^2_\rho} \, = \,
 \frac{E^2_z (\omega ,{\bf k})}{E^2_\rho (\omega ,{\bf k})} \, = \,
 \frac{{\bf q}^2}{k^2}.
\end{equation}
A short calculation leads to the following expression for the differential
spectrum of emitted gluons
\begin{equation}\label{spectrum}
 \frac{1}{\omega } \, \frac{dW}{dz d\omega  d (\tan^2 \theta)} =
 C_V g^2 \frac{\tan^2
 \theta}{1+\tan^2 \theta} \,
 {\cal P} \left[ \tan^2 \theta|\,\zeta(\omega ),\Gamma(\omega ) \right],
\end{equation}
where
\begin{equation}
 {\cal P} \left[ \tan \theta|\,\zeta(\omega ),\Gamma(\omega ) \right] \, =
 \,\frac{1}{\pi} \, \frac{ \tilde \Gamma (\omega )}{(\tan^2 \theta -
 \tilde \zeta (\omega ))^2 + \tilde \Gamma^2 (\omega )}
\end{equation}
and
\begin{eqnarray}
 \tilde \zeta (\omega ) & = & \epsilon_1 (\omega )-1, \nonumber \\
 \tilde \Gamma (\omega ) &  = & \epsilon_2(\omega ).
\end{eqnarray}
Finally, a transformation to the variable $\cos^2 \theta$
leads to the spectrum of Eq.~(\ref{cherloss0})
\cite{G02,GS03}
\begin{equation}
 \frac{1}{\omega } \frac{d W}{dz \,d\omega  \, d \cos \theta} =
 \frac{2 C_V g^2}{\pi}
 \frac{\cos \theta (1-\cos^2\theta) \Gamma(\omega )}
 {\left(\cos^2\theta-\zeta(\omega )\right)^2+\Gamma^2(\omega )}.
\end{equation}


\end{document}